\documentclass[iop]{emulateapj}

\slugcomment{}

\shorttitle{The progressive fragmentation of 332P}
\shortauthors{Kleyna et al.}

\usepackage{natbib}
\usepackage{amsmath}
\usepackage{color}

\begin{document}

\title{The progressive fragmentation of 332P/Ikeya-Murakami}


\author{Kleyna,~J.~T.\altaffilmark{1},
       Ye,~Q.-Z.  \altaffilmark{2},
       Hui,~M.-T. \altaffilmark{3}, \\
       Meech, K.~J. \altaffilmark{1},
       Wainscoat, R. \altaffilmark{1},
       Micheli, M. \altaffilmark{4},
       Keane, J.~V. \altaffilmark{1},
       Weaver, H.~A. \altaffilmark{5},
       Weryk, R. \altaffilmark{1}}


\affil{$^{1}$Institute for Astronomy, University of Hawai`i at Manoa,
  Honolulu, HI, 96822, USA; kleyna@ifa.hawaii.edu}
\affil{$^{2}$Department of Physics and Astronomy,
  The University of Western Ontario, London, Ontario N6A 3K7, Canada}
\affil{$^{3}$Earth, Planetary, and Space Sciences, UCLA,
  Los Angeles, CA 90095}
\affil{$^{4}$ESA SSA-NEO Coordination Centre,  00044 Frascati (RM), Italy}
\affil{$^{5}$The Johns Hopkins University Applied Physics Laboratory,
  Laurel, Maryland, 20723, USA}


\begin{abstract}
  We describe 2016 January to April observations of the fragments of
  332P/Ikeya-Murakami, a comet earlier observed in a 2010 October
  outburst \citep{Ishiguro2014ApJ787}. We present photometry of the
  fragments, and perform simulations to infer the time of breakup.  We
  argue that the eastern-most rapidly brightening fragment ($F4$) best
  corresponds to the original nucleus, rather than the initial bright
  fragment $F1$.  We compute radial and tangential non-gravitational
  parameters, $A_1 = (1.5 \pm 0.4) \times 10^{-8}$ AU day$^{-2}$ and
  $(7.2 \pm 1.9) \times 10^{-9}$ AU day$^{-2}$; both are consistent
  with zero at the $4\sigma$ level.  Monte Carlo simulations indicate
  that the fragments were emitted on the outbound journey well after
  the 2010 outburst, with bright fragment $F1$ splitting in mid--2013
  and the fainter fragments within months of the 2016 January
  recovery.  Western fragment $F7$ is the oldest, dating from 2011.
  We suggest that the delayed onset of the splitting is consistent
  with a self-propagating crystallization of water ice.
  
\end{abstract}





\section{Introduction}
\label{sec:introduction}

 A comet discovered in PanSTARRS1 \citep{Kaiser2002SPIE} survey data
 on 2015 Dec.~31 \citep{Weryk2016CBET} was determined to be the
 recovery of comet 332P/Ikeya-Murakami (previously known as 332P =
 P/2015 Y2), which was discovered in 2010 after a massive outburst
 \citep{Ishiguro2014ApJ787}.  Follow up observations using MegaCam on
 the Canada-France-Hawai`i telescope (CFHT) on 2016 January 1 showed
 another comet $\sim 100\arcsec$ southwest of 332P.  Subsequent
 observations showed the new comet had many pieces and was undergoing
 a fragmentation.  \cite{Sekanina2016CBET4235} suggested that the
 breakup began in 2010 and that what appeared to be the 
 eastern-most nucleus (our $F2$) is the primary;
 \cite{Sekanina2016CBET4250} argued that a new further-east component
 (their $C$, our $F4$)
 \footnote{ Fig.~\ref{fig:photometry} contains a translation of IAU
 fragment designations to our $Fn$ identifiers.} is the primary.

 We obtained data on 12 nights between 1 January and 10 February 2016,
 and identified at least 17 fragments, some of which are shown in
 Fig.~\ref{fig:fragimg}. The situation is very dynamic with rapid
 changes in brightness between fragments and new components appearing
 (fragment $F9$ appeared on January 7).  Both the initial outburst and
 this multiplicity of nuclei are observed near perihelion (True Anomaly
 $TA=11^\circ$ in 2010; $-44^\circ<TA<16^\circ$ here.)

 Comet splitting may be common, but detailed characterization of a
 split nucleus is relatively rare. There have been observations of
 $\sim 45$ split comets in the last 150 years, with only a handful
 well studied \citep{BoenhardtCometsII}.  Splitting is an efficient
 mass loss mechanism for comets, and is important in their
 evolution. Two types of splitting are known: (1) splitting into a few
 pieces, leaving a surviving nucleus; and (2) catastrophic splitting
 into many sub-km fragments. Various mechanisms -- e.g., tides,
 rotation, thermal stress, gas pressure, and impacts -- have been
 proposed for splitting \citep{BoenhardtCometsII}, but only in the
 case of comet Shoemaker-Levy 9 was the mechanism well understood
 (tidal break up, \cite{Sekanina1994AASL9}). Importantly, the size
 distribution and composition the ejected material can be related to
 the internal structure of the comet
 \citep{Belton2015IcarusJupComDis}.

 In this paper, we present photometry and astrometry of these central
 fragments of 332P, and perform dynamical simulations to examine the
 time span over which 332P broke up.  We attempt to use the time of
 the breakup in relation to perihelion to infer the breakup mechanism.



\section{Observations}
\label{sec:observations}

Our principal observations were obtained with MegaCam on the 3.6m CFHT
telescope, on 2016 1 January to 13 April  (Table
\ref{table:observations}).  The two initial nights used the $r$ filter
with exposures of 60s and 90s.  Subsequent nights were observed in the
more sensitive wide-band $gri$ filter for 120s to 180s.  Additionally,
 we obtained $R_C$-band observations with the UH2.2m and
Tek2048, on UT 16 Jan 2015 in the middle of the gap between CFHT runs.

\begin{deluxetable*}{lllllllllll} 
\tablewidth{0pt}
\tablecaption{Observations \label{table:observations}}
\tablehead{
\colhead{UT Date}   & 
\colhead{UT Times}   & 
\colhead{Telescope/Instrument} &
\colhead{Filter}      &
\colhead{FWHM  \tablenotemark{a}}      &
\colhead{No.~exp.\tablenotemark{b}}    &  
\colhead{Exp.~time\tablenotemark{c}} &
\colhead{$r_h$\tablenotemark{d}}  & 
\colhead{$\Delta$ \tablenotemark{e}} & 
\colhead{$\alpha$ \tablenotemark{f}} & 
\colhead{$TA$ \tablenotemark{g}}  
}
\startdata
2016-01-01 & 11:36--11:40 & CFHT/MegaCam & $r$ & 0.78 & 3 & 60 & 1.73 & 0.82 & 18.68 & -43.36\\
2016-01-02 & 11:30--11:35 & CFHT/MegaCam & $r$ & 0.74 & 3 & 90 & 1.72 & 0.82 & 18.34 & -42.86\\
2016-01-03 & 13:39--13:44 & CFHT/MegaCam & $gri$ & 0.69 & 3 & 120 & 1.72 & 0.81 & 17.96 & -42.30\\
2016-01-05 & 13:00--13:16 & CFHT/MegaCam & $gri$ & 1.12 & 5 & 180 & 1.71 & 0.79 & 17.27 & -41.29\\
2016-01-06 & 12:50--13:09 & CFHT/MegaCam & $gri$ & 0.75 & 6 & 180 & 1.71 & 0.79 & 16.92 & -40.78\\
2016-01-07 & 13:38--13:46 & CFHT/MegaCam & $gri$ & 0.90 & 3 & 180 & 1.71 & 0.78 & 16.56 & -40.25\\
2016-01-08 & 10:06--13:14 & CFHT/MegaCam & $gri$ & 0.91 & 12 & 180 & 1.70 & 0.77 & 16.24 & -39.77\\
2016-01-16 & 08:53--12:01 & UH2.2m/Tek2048 & $R_C$ & 0.93 & 5 & 600 & 1.68 & 0.73 & 13.58 & -35.57\\
2016-01-30 & 09:27--09:35 & CFHT/MegaCam & $gri$ & 1.53 & 3 & 180 & 1.64 & 0.67 & 11.43 & -27.87\\
2016-01-31 & 11:47--11:55 & CFHT/MegaCam & $gri$ & 1.50 & 3 & 180 & 1.63 & 0.67 & 11.50 & -27.25\\
2016-02-05 & 09:58--10:06 & CFHT/MegaCam & $gri$ & 1.51 & 3 & 180 & 1.62 & 0.66 & 12.33 & -24.44\\
2016-02-10 & 10:15--10:23 & CFHT/MegaCam & $gri$ & 1.72 & 3 & 180 & 1.61 & 0.65 & 13.86 & -21.54\\
2016-02-11 & 10:37--10:45 & CFHT/MegaCam & $gri$ & 1.63 & 3 & 180 & 1.61 & 0.65 & 14.23 & -20.95\\
2016-02-12 & 10:05--10:13 & CFHT/MegaCam & $gri$ & 1.69 & 3 & 180 & 1.61 & 0.65 & 14.60 & -20.38\\
2016-02-13 & 10:27--10:34 & CFHT/MegaCam & $gri$ & 0.73 & 3 & 180 & 1.60 & 0.65 & 15.00 & -19.78\\
2016-02-14 & 06:15--06:23 & CFHT/MegaCam & $gri$ & 1.58 & 3 & 180 & 1.60 & 0.65 & 15.34 & -19.30\\
2016-03-31 & 06:46--06:54 & CFHT/MegaCam & $gri$ & 0.70 & 3 & 180 & 1.58 & 0.81 & 32.73 & 8.51\\
2016-04-05 & 05:39--05:47 & CFHT/MegaCam & $gri$ & 0.73 & 3 & 180 & 1.58 & 0.84 & 33.76 & 11.50\\
2016-04-09 & 05:36--05:43 & CFHT/MegaCam & $gri$ & 0.69 & 3 & 180 & 1.59 & 0.87 & 34.47 & 13.90\\
2016-04-13 & 06:07--06:26 & CFHT/MegaCam & $gri$ & 0.70 & 6 & 180 & 1.59 & 0.90 & 35.07 & 16.30\\
\enddata 
\tablenotetext{a}{Median point source FWHM in arcsec.} 
\tablenotetext{b}{Number of exposures.}
\tablenotetext{c}{Exposure time, in seconds.}
\tablenotetext{d}{Heliocentric distance, AU.} 
\tablenotetext{e}{Geocentric distance, AU.}
\tablenotetext{f}{Phase angle, degrees.} 
\tablenotetext{g}{True anomaly, degrees.}
\end{deluxetable*}

Conditions were good on all nights, with little variation in the
photometric zero point, and seeing FWHM ranged from 0.7\arcsec\ to 1.7\arcsec.

\begin{figure*}
   \epsscale{1.0}
   \plotone{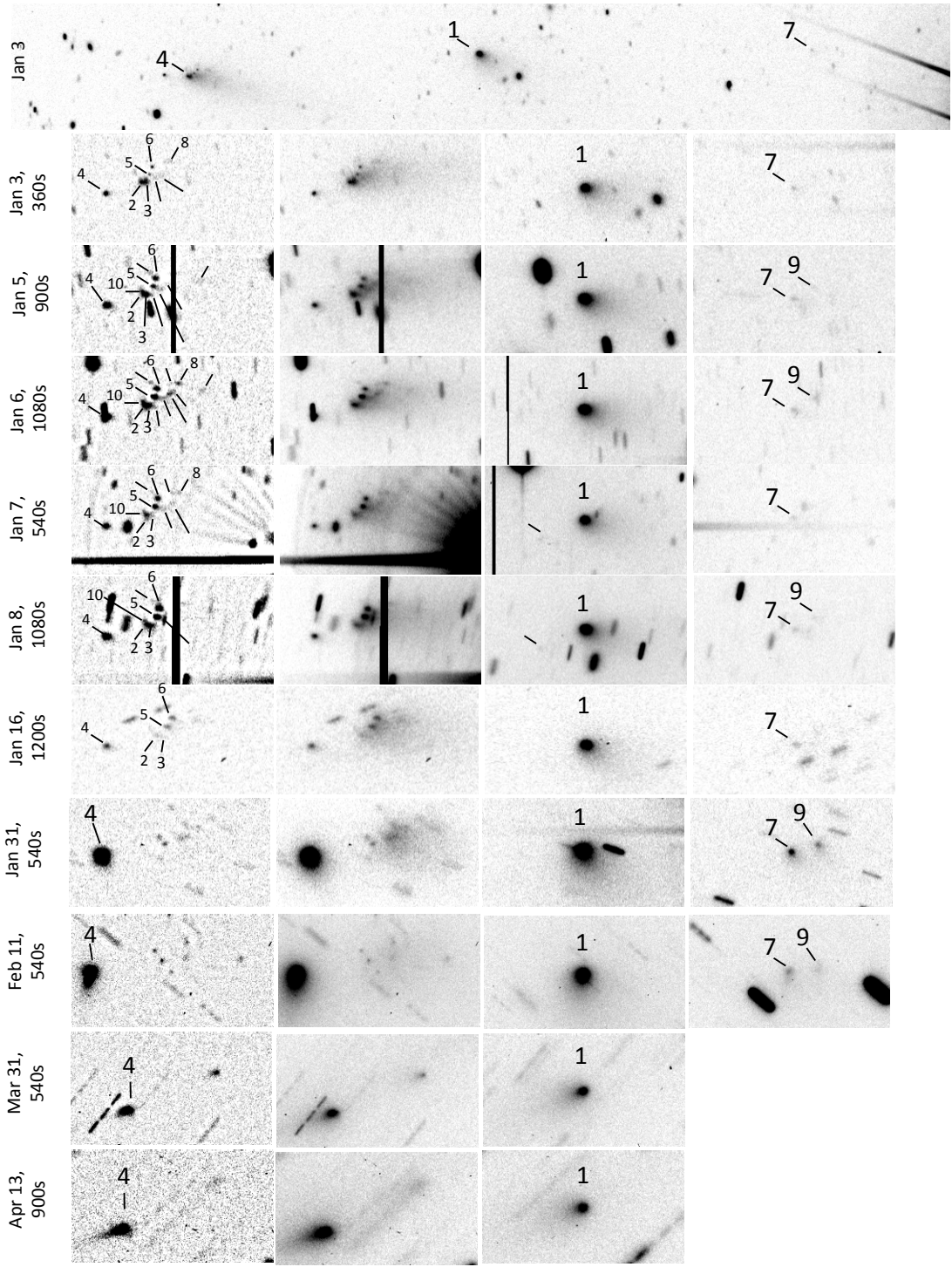}
   \caption{Montage of images of 332P from 2016 January 3 to
     13 Apr.  The top and bottom strips show a portion of the full
     image rotated by $21^\circ$. Each other (multi--pane) row shows
     $45\arcsec \times 25\arcsec$ regions centered on fragments $F4$,
     $F1$, and $F7$; N is up and E is to the left.  The left column is
     an enhanced (unsharp mask) view of $F4$ to show the fragments.
     Un-numbered fragments are detected only over a short span of time,
     and this figure does not show fragments (such as 332P-G) that are
     far from F1 and F4.
     \label{fig:fragimg}} 
 \end{figure*}

Figure \ref{fig:fragimg} shows the rapid evolution of the sub-nuclei in
relative and absolute brightness, position, and distinctness.

We found additional $i$ and $r$-band images in the Pan-STARRS1 (PS1)
archive from 2015 Oct 10 and 2015 December 4.  No evidence of any fragments
is seen in October, but fragment $F1$ and the $F2$ through $F6$ cluster
are seen in December.  Because these are short 45s images on a small
1.8m telescope, it was necessary to stack four images to see the
December PS1 $r$-band detections.

\section{Photometry}
\label{sec:photometry}

The MegaCam images were bias--subtracted and flattened by the Elixir
pipeline at CFHT.  We fit a world coordinate system (WCS) using the
AstrOmatic {\sc SWarp} tool \citep{Bertin2002Terapix}, and used our
custom calibration pipeline to give an absolute magnitude calibration
using SDSS DR8 \citep{SDSSDR82011}.  For the $r$ filter, we ignored
the minor 1\% difference between SDSS $r$ and MegaCam $r$.  For the
broadband $gri$ filter, we used a transformation to a $gri$ AB
magnitude computed by Simon Prunet (personal communication), where
\begin{multline}
  gri_{AB}=-2.5 \log_{10} [ 0.4388 \times 10^{-0.4 g} \\    + 0.4146 \times 10^{-0.4 r} + 0.1490\times 10^{-0.4 i} ]
\end{multline}
and $g,r$ and $i$ are SDSS AB magnitudes.  Using measurements of
fragments $F1$ and $F4$ in a pair of images on 30 Jan, we convert $r$ to
$gri$ as $gri=r+0.18$, and present all photometry.

Because the fragments were crowded, diffuse, and (for $F2$ and $F3$)
overlapping, we applied a simple aperture correction to compute
individual magnitudes. We computed the magnitude of isolated fragment
$F1$ in each image in a large 5\arcsec\ diameter aperture, and used this
to correct the magnitude of all measured fragments to a smaller
2\arcsec\ aperture.  Additionally, we increased the size of this
calibration aperture inversely with geocentric distance, to maintain a
constant physical aperture size.  The initially brightest fragment, $F1$, had a
magnitude of $r\sim gri \sim 20$, and the faintest fragment $F9$ had
$gri\sim 25$.

\begin{figure}
  \epsscale{1.15}
  \plotone{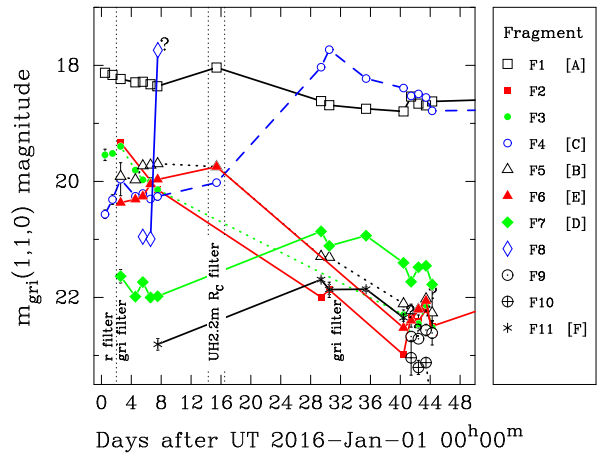}
  \caption{First 50 days of fragment photometry, magnitudes
    $m(1,1,0)$, with a question mark indicating a data point based on
    a single measurement and near a bright star diffraction spike. $r$
    magnitudes of first two nights are converted to $gri$.  With a few
    exceptions, the error bars are smaller than the graph symbols,
    omitting systematic uncertainties in the phase correction.  The
    letters in the square brackets in the legend indicate the current
    CBET and JPL Horizons designations of the fragments; for example,
    our $F1$ corresponds to 332P-A.\label{fig:photometry}. }
\end{figure}

Figure \ref{fig:photometry} shows the absolute photometry $m(1,1,0)$
obtained by adjusting calibrated magnitude $m$ to geocentric and
heliocentric distances $\Delta=r_h=1\rm\,AU$ and to phase angle $\alpha=0^\circ$,
with the cometary dust phase coefficient $\beta=0.02$ derived from
\cite{Meech1987AA} and \cite{Krasnopolsky1987AA}; however \cite{Marcus2007} 
gives a slightly larger empirical value of $\beta=0.031$.
\begin{equation}
  m(1,1,0) = m - 5 \log_{10} \left( r_h \Delta \right) - 0.02 \times \alpha
\end{equation}  The uncertainties shown by the error bars are computed
from the scatter of measurements taken during one night.
 
It is immediately evident that the brightest fragment $F1$ is growing
fainter as it approaches perihelion, a trend that is followed by
$F2,F3$, and faint fragment $F7$.  However, fragment $F4$ appears to
rise in brightness in the initial $r$-band measurements, and $F5$ and
$F6$ are growing steadily brighter, suggesting that the cloud is
evolving rapidly.  Most notably, by 30 January, $F4$ has increased in
brightness by two magnitudes, and is clearly the brightest object.  On
31 January, its $m(1,0,0)\approx 17.5$ is only about one magnitude
fainter than the fading value after the 2010 outburst \citep[][Table
  2]{Ishiguro2014ApJ787}.

In addition to the data in Fig.~\ref{fig:photometry}, we performed
photometry on the 4 December PS1 images. We measured both fragment $F1$
and the diffuse assemblage at $F2$ to have identical magnitudes
$r=21.9\pm 0.2$ and $m(1,1,0)_{gri}=17.7\pm 0.2$.  For $F1$ this
magnitude is consistent with the dimming trend seen in the subsequent
CFHT observations.  For $F2$, this value is roughly consistent with the
summed flux of $F2,F3,F5,F6$ in the CFHT observations suggesting little
evolution in $F1$ or $F2,F3,F5,F6$ from 4 Dec.  The surprising result was
the complete absence of concentrated fragment $F4$ in the PS1 4
December image. The absence of $F4$ may be consistent with the ongoing
brightening as seen in Fig.~\ref{fig:photometry}, but may also arise
from the depth limitations of the PS1 data.

In a supplementary table to this paper, we present astrometry and
photometry of the fragments in Fig.~\ref{fig:photometry} extending
to 13 April, omitting the low quality PS1 data.  The full data set in
the supplementary data table shows fragments F1 and F4 brightening
slightly, and F2 brightening by $\sim 2$ mag.

\section{Non-gravitational Forces from Astrometry}
\label{sec:nongrav}


The radial and transverse nongravitational parameters, respectively
$A_1$ and $A_2$, from a symmetric nongravitational force model devised
by Marsden et al. (1973), of component F4 were obtained by Hui et al.
(in preparation) using EXORB8{\footnote{\tt
{http://www.solexorb.it/Solex120/Download.html}}} to be $A_1 =
(1.5 \pm 0.4) \times 10^{-8}$ AU day$^{-2}$ and $(7.2 \pm 1.9) \times
10^{-9}$ AU day$^{-2}$, taking measurements and astrometry from the
Minor Planet Center through UT 2016 April 13, confirming existence of
a nongravitational force.  $A_2>0$ is consistent with prograde
rotation \citep{YeomansCometsII}.  Admittedly the nongravitational
model may not be perfectly suitable to the case of component F4, as a
small but clear systematic bias trend in astrometric residuals
starting from 2016 April 09 is identified by Hui et al. However,
taking the complexity of the cascading fragmentation experienced by
this comet into consideration, and the short observed arc of the 2016
apparition, we think that the application of the nongravitational
model is still meaningful.

We also attempted to prove that component F4 is the primary component,
rather than the earliest detected component F1. If the identification
of F4 as the primary is incorrect, we would expect to see a larger RMS
in the orbital solution compared to a correct linkage. By following
the same methods and procedures, we set to establish a linked orbit
between observations from 2010-2011 and the current observations of
component F1. The effort turned out to be very successful and the
solution quickly converged after only a few iterations. Indeed, the
RMS of the best fit is even slightly better ($\sim$0\arcsec.1) than
that of component F4, regardless of ballistic or nongravitational
solutions. However, this does not necessarily mean that the
identification of component F4 as the primary of comet 332P is
erroneous, because the positional uncertainties of F4 are generally
greater than those of F1, as a consequence of the well-defined
morphology of component F1, and its isolation from any other fragments
of 332P.  The separation speeds of fragments split off cometary nuclei
are generally found to be $\sim$0.1 -- 1 m s$^{-1}$ (Sekanina 1982),
which is hard to detect by fitting an impulse from optical astrometric
observations alone (Farnocchia, private communication). Therefore the
successful linkage between the 2010-2011 observations and observations
of component F1 of the current apparition is not surprising. In order
to determine the genetic relationship between components F1 and F4,
alternative dynamic methods will have to be used.

\section{Fragment dynamics}
\label{sec:dynamics}

To understand the dynamical properties of the fragments, we make use
of the Monte Carlo model developed in our earlier works
\citep[e.g.][]{Ye2014c,Ye2015i,Ye2016a}. In detail, our approach
creates an ensemble of fragment ejection directions, velocities, and
times, and finds those emission events that best match the final
observations. Intervals of ejection time that produce a larger number
of viable matches with the data are deemed more likely than intervals
that produce fewer matches.

The continuing fragmentation of the comet makes it challenging to follow each 
fragment over an extended period of time, especially for fragments in the 
debris field surrounding fragment $F3$. Hence, we focus on four distinct 
fragments $F1$, $F3$, $F4$ and $F7$ only. Component $F4$ is the likely primary
as discussed; fragments $F1$ and $F7$ are either very bright or distant from 
the remainder, allowing robust night-to-night associations; fragment $F3$ 
began as the brightest fragment in the debris field that was unambiguously 
followed for the longest time span (2016 Jan. 1 to 8), making 
it a good representative of the debris field as a whole. All other mini
fragments in the debris field are extremely diffuse and it is impossible to
uniquely track them for more than a few days.

Unarguably, all fragments are the descendants from the primary component $F4$. 
The question is whether all fragments are the \textit{direct} descendants of 
$F4$. It is plausible that the fragments continue to split into smaller 
fragments as already been hinted by the $F2$-$F3$ debris field. For a given 
fragment, any fragment to its east could be its direct parent, i.e. $F7$ could 
be the direct descendant of either $F1$, $F3$ or $F4$, and $F1$ could be
directly from either $F3$ or $F4$. However, both $F1$ and $F7$ are apparently
``healthy'' fragments while $F3$ and its immediate companions are only visible 
intermittently, we feel it is unlikely for $F3$ (or its direct progenitor) to
be the parent of either $F1$ or $F7$. Hence, we explore these two scenarios:

\begin{enumerate}
 \item $F4$ as the parent of $F1$, $F3$ and $F7$; and
 \item $F4$ as the parent of $F1$ and $F3$, while $F1$ as the parent of $F7$.
\end{enumerate}

A total of $\sim15$ million virtual fragments are released in a 10-day
step from the 2010 outburst (circa. 2010 Nov.~1) to the recovery in 2016 (circa.
2016 Jan.~1), from the parent component (either $F1$ or $F4$). The fragments
are ejected isotropically at random speeds between 0 and $1~\mathrm{m~s^{-1}}$,
a range determined by \citet{Sekanina1982Comets}. The dynamics of the fragments are 
then mainly determined by solar gravity as well as the net radial force due to 
anisotropic outgassing of the respective fragment, described by the ratio 
between the force and solar gravity, $\gamma$ \citep[c.f.][]{Sekanina1977f}. We 
then integrate the fragments to the epoch of 2016 Jan. 8.0 TT using the 15th 
order RADAU integrator embedded the MERCURY6 package \citep{Everhart1985, 
Chambers1999d}, accounting for the gravitational perturbations from the eight 
major planets (including the Earth-Moon barycenter), and radiation pressure. At 
the end, we compare the modeled \textit{relative} positions between the primary 
and the fragment with the observations, and record the solutions when the 
observed minus calculated ($O-C$) error is less than $1\arcsec$ (roughly 
one FWHM). Henceforth, we refer this set of solutions as ``good'' solutions.

\begin{figure*}[h]
\includegraphics[width=\textwidth]{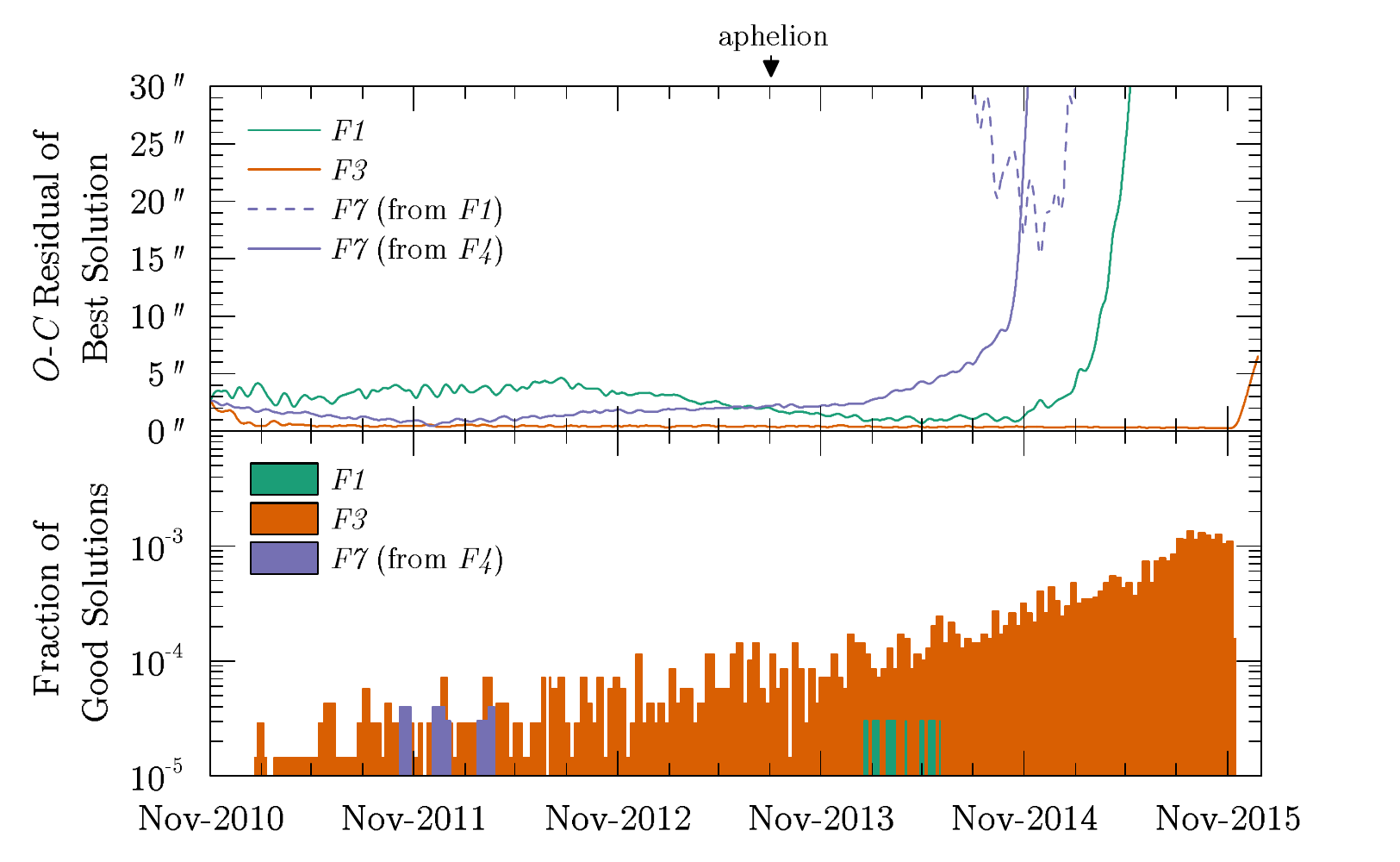}
\caption{Upper panel: best solutions (i.e. solutions with smallest 
$O-C$ residuals) at different split epochs; bottom panel: fraction of ``good'' 
solutions (i.e. solutions with $O-C$ residual less than half FWHM or 
$0.5''$). Only those regions of the bottom panel that produce the largest 
fraction of good detections represent viable ejection times. For example, $F1$ 
(green) could only have been ejected in early to mid-2014.}
\label{fig:mc-min}
\end{figure*}

\begin{figure*}[h]
\includegraphics[width=\textwidth]{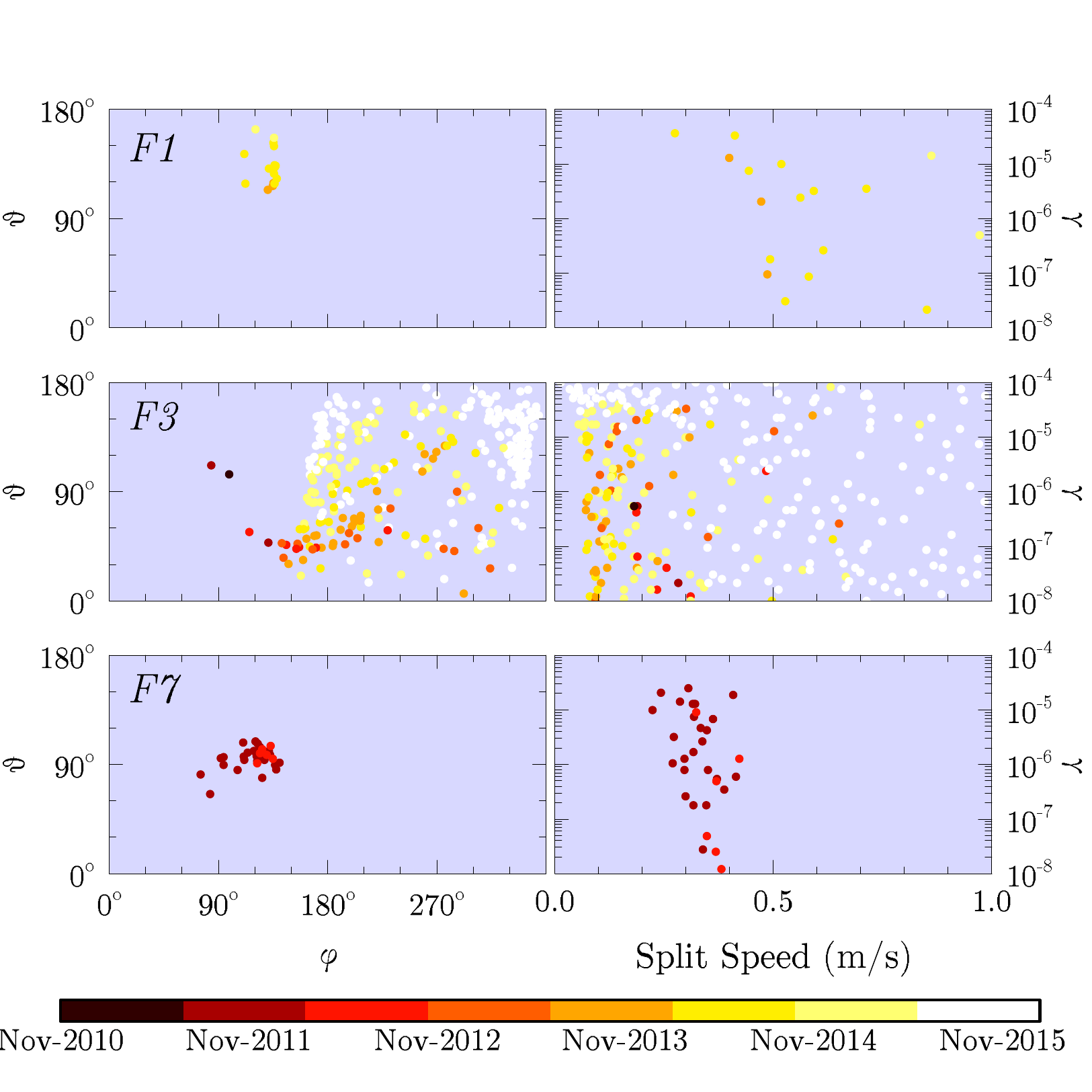}
\caption{Split direction (left column) as well as split speed and deceleration 
parameter $\gamma$ (right column) for good solutions ($O-C$ residuals less than 
$0.5''$) for fragments $F1$ (upper row), $F3$ (middle row) and $F7$ (bottom
row). The data points for fragment $F3$ are under-sampled for clarity. Color
denotes different split epoch, filled circle denotes to the scenario that
fragment $F4$ is the parent and cross denotes to the scenario that fragment $F1$
is the parent. $\varphi$ and $\vartheta$ are the longitude and latitude in a
standard spherical coordinate system, with the $Z$ axis pointing toward the sun,
the orthogonal $X$ axis aligned in the negative orbit direction, and the $Y$
axis pointing north from the $XZ$ orbital plane. $\gamma$ is the ratio of 
anisotropic outgassing force of the respective fragment to the solar gravity.}
\label{fig:mc-best}
\end{figure*}

The demographics of the ``good'' solutions are shown as
Figure~\ref{fig:mc-min} and~\ref{fig:mc-best}.  Fragment direction is
expressed in the comet's reference frame, with the $Z$ axis pointing
toward the sun, the orthogonal $X$ axis aligned in the negative orbit
direction, and the $Y$ axis pointing north from the $XZ$ orbital
plane.  A fragment's ejection direction perpendicular to an assumed
spherical surface is specified by standard spherical longitude
$\varphi$ and latitude $\vartheta$.  That is,
$x=\sin{\vartheta}\cos{\varphi}$, $y=\sin{\vartheta}\sin{\varphi}$,
and $z=\cos{\vartheta}$.  Then $\vartheta=0^\circ$ points at the sun,
$\varphi=0^\circ$ points in the negative sense of orbital motion, and
$\varphi=90^\circ$ points north.

We immediately note that the solutions do not congregate around or shortly after
the 2010 outburst, but instead mostly scatter along the comet's outbound 
journey. This indicates that the surviving fragments are not immediately 
produced by the major outburst. For each fragment:

\begin{enumerate}
 \item The separation of $F1$ from the main nucleus most likely occurred 
in early 2014, with an uncertainty of about 6 months. The median 
solution is $\vartheta=120^\circ$, $\varphi=130^\circ$, split speed 
$0.5\pm0.1~\mathrm{m~s^{-1}}$, and $\gamma \sim 10^{-6}$, representing 
an emission in the northern hemisphere at the night-side.
   
 \item The split of fragment $F3$ is complicated by the fact that $F3$ is 
observationally less constrained. However, we note that most good solutions 
congregate over recent epochs, within a few months from the recovery in 
late 2015, consistent with the fact that fragment $F3$ is much closer to 
the primary than fragment $F1$.  This, in turn, implies that the split between 
$F3$ and the primary occur much later than the split between $F1$ and the 
primary. The good solutions are very scattered across the parameter spaces, 
but do seem to congregate in the nucleus' southern hemisphere at the 
night-side (i.e. $180^\circ<\varphi<360^\circ$, $90^\circ<\vartheta<180^\circ$).
   
 \item The simulation favors $F4$ as the direct parent of fragment $F7$, as 
the solution to $F1$ leaves a large systematic residual. The split epoch is 
around late 2011 to early 2012. The median good solutions are near 
$\vartheta=90^\circ$, $\varphi=120^\circ$, or in the twilight zone of the 
southern hemisphere, with a split speed of $\sim 0.3~\mathrm{m~s^{-1}}$) and
a $\gamma$ of $\sim10^{-6}$.
\end{enumerate}

In general, our solution agrees with the general conclusion by 
\citet{Sekanina2016a, Sekanina2016b}, that all the surviving fragments are 
likely not immediately produced by the 2010 outburst. Further refinement of 
split parameters is challenging, as the determination is highly sensitive to 
the quality of the astrometry. This is especially true for the sub-fragments in 
the debris field which cross-identification between observations from different 
night/observatory needs to be done with great care.

\section{Discussion}

Initially, fragment $F1$ was brighter than the nucleus
$F4$.  However, as noted by \cite{Sekanina1982Comets,Sekanina2009ICQ},
split comet fragment size is often unrelated to dust flux.

The previous far larger outburst of comet 17P/Holmes also produced
fragments, but these were seen immediately after the event
\citep{Stevenson2010AJ}, in contrast to the 2010 outburst of P/2010
V1.  In both events, a proposed mechanism for the outburst was the
runaway exothermic conversion of CO-laden amorphous ice to the
crystalline form, triggered by a perihelion heat wave from the surface
\citep{Sekanina2009ICQ,Ishiguro2014ApJ787}.  It is possible that the
delay in fragment emission from the initial outburst is the result of
a smoldering ongoing crystallization process initiated by the 2010
perihelion.  Fig.~1 of \cite{Sekanina2009ICQ} shows that the
crystallization process has a many-year slow-growth phase before
accelerating catastrophically, in agreement with the delayed
fragmentation we observe. If such a process hit pockets of CO rich
ice, or if it propagated into deep reservoirs where it reached a
critical temperature of $\sim 130\, K$ months or years later, it could
create a sequence of delayed post-perihelion splitting events.

In contrast to amorphous ice crystallization, the direct sublimation of
CO or other volatiles, even if they survived at the heliocentric
distance of 332P, would be a self-quenching endothermic process.

\section{Conclusions}

We performed photometric analysis and dynamical simulations of the
central components of fragmented comet 332P. Dynamical arguments
indicate that $F4$ corresponds to the primary fragment, which is
within a magnitude of the $m(1,1,0)$ of the fading post-outburst
nucleus in 2011.

Dynamical simulations indicate that the breakup occurred over a period
of years after the 2010 October outburst, with eastern fragments $F1$ and
$F7$ breaking off as early as 2011.

\acknowledgments

We thank Simon Prunet for the transformation from $g,r,i$ to the
MegaCam $gri$ filter.  Q.-Z. thanks Peter Brown for support as well as
Paul Wiegert for computational resources. Part of the numerical
simulations were conducted using the facilities of the Shared
Hierarchical Academic Research Computing Network
(SHARCNET:www.sharcnet.ca) and Compute/Calcul Canada.  We thank Dina
Prialnik for useful discussions of the physics of activity, and Eva
Lilly for UH2.2m observations. Part of the numerical simulations were
conducted using the facilities of the Shared Hierarchical Academic
Research Computing Network (SHARCNET:www.sharcnet.ca) and
Compute/Calcul Canada.  M.-T. thanks Aldo Vitagliano for an improved
version of EXORB.  RJW acknowledges support by NASA under grants
NNX12AR65G and NNX14AM74G.  KJM, JK and JVK acknowledge support
through the NASA Astrobiology Institute under Cooperative Agreement
NNA08DA77A, and partial support through an award from the National
Science Foundation AST1413736.  MTH is supported by a NASA Solar
Systems Observations program to D.~Jewitt. Observations were obtained
with MegaPrime/MegaCam, a joint project of CFHT and CEA/DAPNIA, at the
Canada-France-Hawaii Telescope (CFHT).

{\it Facilities:} \facility{CFHT (MegaCam)}, \facility{Pan-STARRS1},
\facility{UH 2.2m}


\end{document}